\documentclass[12pt,dvips]{article}
\usepackage{epsfig}
\usepackage{float}
\newcommand{\bea}{\begin{eqnarray}}
\newcommand{\eea}{\end{eqnarray}}

\newcommand{\gsim}
{{\;\raise0.3ex\hbox{$>$\kern-0.75em\raise-1.1ex\hbox{$\sim$}}\;}}
\newcommand{\lsim}
{{\;\raise0.3ex\hbox{$<$\kern-0.75em\raise-1.1ex\hbox{$\sim$}}\;}}

\begin{document}
\begin{flushright}
{ROME1-1389-2004} \\
\end{flushright}

\begin{center}

{\Large\bf SUSY Resonances from UHE neutralinos \\
\vskip 0.3cm
in Neutrino Telescopes and in the Sky
}\\[10mm]
Anindya Datta,$\;$
Daniele Fargion,$\;$
and $\,$ Barbara Mele $\,$\footnote{E-mail: 
Anindya.Datta@roma1.infn.it, Daniele.Fargion@roma1.infn.it,
Barbara.Mele@roma1.infn.it} \\

\vskip 0.5cm
{\em INFN, Sezione di Roma, and \\ Dip. di Fisica, Universit\`a
La Sapienza, P. le A. Moro 2, I-00185, Rome, Italy.}
\end{center}

\vskip 20pt
\begin{abstract}
\noindent
In the {\it Top-down} scenarios, the decay of super-heavy particles ($m
\sim 10^{12-16} $GeV), 
situated in dark-matter halos not very far from
our Galaxy, can explain the ultra-high-energy (UHE) cosmic-ray
spectrum beyond the Griesen-Zatasepin-Kuzmin  cut-off. In case
the dynamics of this decay is governed by the minimal supersymmetric
standard model, a major component of the UHE cosmic-ray flux at
PeV-EeV energies could be given by the lightest {\it neutralino}
$\chi^0_1$, that is the lightest stable supersymmetric particle. Then,
the signal of UHE $\chi^0_1$'s on earth might emerge over the
interactions of a comparable neutrino component. We compute the event
rates for the resonant production of {\it right selectrons}
($\tilde{e}_R$) and {\it right squarks} ($\tilde{q}_R$) in mSUGRA,
when UHE neutralinos of energy $E_{\chi}\gsim 10^5$ GeV scatter off
electrons and quarks in an earth-based detector like IceCube. 
When the resonant channel dominates in the total $\chi^0_1e,\chi^0_1q$
scattering cross section, the only model parameters affecting
the corresponding visible signal rates 
turn out to be the  physical masses of the resonant right-scalar 
and of the lightest neutralino.
We compare the expected number of supersymmetric events with
the rates corresponding to the expected Glashow $W$ resonance 
and to the continuum
UHE $\nu N$ scattering for realistic power-law spectra. We find that
the event rate in the leptonic selectron
channel is particularly promising, and can reach a few tens for a one-year
exposure in IceCube.  Finally, we note that UHE neutralinos at much
higher energies (up to hundreds ZeV) may produce {\it sneutrino}
($\tilde{\nu}$) resonances by scattering off relic neutrinos in the
Local Group hot dark halo. The consequent $\tilde{\nu}$ {\it burst} into
hadronic final states could mimic  {\it Z-burst} events, although with 
quite smaller conversion efficiency.
\end{abstract}

\section{Introduction}

The presence of a ultra-high-energy (UHE) component in the cosmic-ray
spectra beyond the Griesen-Zatasepin-Kuzmin cutoff (GZK) \cite{gzk}, 
as revealed initially by the
experiment Fly's Eye \cite{flys_eye}, subsequently by AGASA
\cite{agasa}, and marginally by Hires \cite{hires}, is presently an
open problem in high-energy astrophysics. The highest-energy particles
in the cosmic-ray spectrum are apparently protons rather than photons,
because of their characteristic hadronic-shower fluorescence light
shapes. Beyond the GZK energies, protons, slowed down by the cosmic
Black Body Radiation, have energy-loss length well below 50 Mpc. This
implies that possible sources of these extremely-high-energy particles
should be within our Galaxy, or inside our Local Group of Galaxies or
nearby Clusters. However, there are no such powerful sources in Milky
Way or nearby Virgo Group, that are correlated with the arrival map of
the UHE cosmic rays (UHECR). There is a timid, and yet unconfirmed,
clustering of AGASA data along the Super-Galactic Plane, not
sufficient to probe the UHECR confinement in observable GZK volumes.
On the contrary, UHECR events exhibit isotropy, a signature
reminiscent of most distant cosmic edges. It is not clear whether the
UHECR spectrum is really showing a GZK cut-off \cite{agasa,hires}. In
order to make our UHECR understanding even more confused, there are
now first evidences of clustered UHECR events by AGASA correlated with
known BL Lacs sources \cite{Gorbunov02}. This UHECR-BL Lacs connection
calls for the UHE Blazing Jet of active galactic nuclei (AGN) as a
source of particle acceleration. Those Bl Lacs sources are mostly at
distances above GZK lengths.

Although the GZK puzzle is still experimentally
not completely settled, in the past and
also recently, there have been quite a few suggestions
 about possible sources and mechanisms
for accelerating particles to such extreme energies
\cite{suggestn,flux_rev}. \\
A class of proposals, generically called {\it Top-down} scenarios,
tries to explain the high-energy part of the spectrum through the
decay of super-massive particles ($m \sim 10^{12-16} \rm{GeV}$)
\cite{massive,ber,susy_oth,toldra,drees}, or by their eventual binary
collapse and annihilation in place of time-tuned
decays \cite{Farg-Khlop04}. 
These particles are generally predicted in Grand Unified
Theories (GUT). Their decay lifetime should be comparable with
the present age of the Universe, so that their decay products
would give rise today to the highest energy part of the cosmic-ray
spectrum. 
Also topological defect decays can give rise to the UHECR primaries
\cite{suggestn}.
 In any case, the decay dynamics of the
super-heavy particles is controlled by the model assumed for
particle interactions. 
In the standard model (SM), the decay after
fragmentation finally results into photons, neutrinos and hadrons.
In supersymmetry (SUSY), UHE secondary stable neutralinos ($\chi^0_1$)
may be born, too \cite{ber,susy_oth,toldra,drees}.

 The minimal supersymmetric
standard model (MSSM) is a well motivated extension of the
standard model of elementary particles \cite{Haber:1984rc}. SUSY
stabilizes the Higgs-boson mass under radiative corrections by
introducing an entire spectrum of supersymmetric partners for the
SM particles. The MSSM  predicts a stable weakly-interacting
heavy neutral particle, called neutralino ($\chi^0_1$), which can
be also a good candidate for cold dark matter\footnote{In general, in
the MSSM, there are 4 neutralinos. Here, we are talking about the
lightest one $\chi^0_1$.}.  If the masses of supersymmetric
particles are at the TeV scale (as needed to stabilize the
Higgs-boson mass), we expect to discover them in a few years at
the CERN LHC \cite{lhc}. In the case  SUSY is the correct theory
of particle interactions, it will govern also the decays of
super-heavy particles. In \cite{susy_oth,toldra,drees}, the
super-heavy particle decays in the MSSM extension of the SM have
been investigated in detail. These decays would involve the entire
spectrum of SUSY partners. At the end of the decay chain, one would be
left with a spectrum (calculable in the MSSM) of all the  stable
particles, that is SM particles plus  stable neutralinos.
 Hence, a crucial prediction of
these models is the presence of neutralinos in the UHECR spectrum
produced by super-heavy objects.

\vskip 0.2cm
Another  possible explanation for the GZK paradox considers relic
light neutrinos as a calorimeter in resonance with incoming UHE
(ZeVs) neutrinos.
The latter may come freely from distances above GZK from
far BL Lacs (hence explaining the UHECR-BL Lacs correlations) or gamma ray
bursts (GRB). They may also arise from topological defects spread
all over dark halos (Galactic or extra-galactic).
 This model, the so-called
{\it Z-burst} model \cite{zburst}  will be  discussed in the following in view
of a new and  analogous ${\it sneutrino}$-burst model.

 
\vskip 0.2cm
According to the discussion above, we shall
assume in this study a rather  
hard neutralino spectrum (i.e. $\sim E_{\chi}^{-1.5}$) that should be typical 
for  the GUT-relics fragmentation. 
In order to estimate the effects of model dependence in 
the spectral index, we will also compare our results
with the ones corresponding to 
a softer {\it Fermi-like} flat spectrum,  
going as $\sim E_{\chi}^{-2}$.
We shall not consider a much harder {\it Z-burst}$-$like
spectrum  ($\sim E_{\chi}^{-1}$) that is usually used in different context,
although such a spectrum could be of relevance for a 
${\it sneutrino}$-burst model.

\vskip 0.2cm
One of the major goals for the upcoming UHECR experiments 
(like Amanda, IceCube, Antares, Nestor) is to
verify these predictions. In \cite{pnath}, the possible
signatures of up-going UHE neutralinos in the future satellite-based 
detector EUSO \cite{euso} have been investigated. The resonant
production of quite heavy squarks ($m \sim 0.8 - 1.2 ~\rm TeV$) in
the $\chi_1^0$ scattering off nucleons in the earth's atmosphere,
after crossing all the earth volume, has been considered. Heavy
squarks are selected so that the neutralino interaction cross
sections are moderate, and the neutralino flux is not completely
stopped by the earth screen (while the corresponding neutrino
background does not survive the passage through the earth). An
analogous process leading to a smaller neutrino earth opacity 
 has been also considered
earlier,  while discussing the up-going $\tau$ neutrinos
\cite{Fargion2002a}.
Note that, in EUSO, the UHE neutralino interaction in the atmosphere is
polluted by more abundant up-going-horizontal $\tau$ air-showers
induced by UHE $\nu_{\tau}$ earth skimming in the terrestrial crust 
\cite{skim_tau}.

\vskip 0.2cm
In the present article, we will focus on the interaction of the
highest energy neutralinos with both electrons and nucleons  in a
earth-based detector like IceCube \cite{icecube}. In particular,
we will study the possibility to detect  the resonant production
of the SUSY partners of electrons and quarks, when their
 masses are larger than  the present experimental limits
derived from direct searches at high-energy colliders ($m_{\tilde
e}\gsim 100$ GeV for selectrons \cite{emass}, 
$m_{\tilde q}\gsim 300$ GeV for squarks\footnote{Even 
lighter squarks are allowed for $m_{\tilde g}\gg m_{\tilde q}$.}
\cite{tevatron}). We will also take into account the constraints on 
the supersymmetry parameter space coming from bounds on the relic density
of the lightest neutralino, in case the latter is responsible for most of 
the cold dark matter in the universe.

Neutralinos with appropriate energy can produce a right/left
selectron ($\tilde e_R$/$\tilde e_L$) resonance in the detector
via the scattering off the detector electrons
$$
\chi_1^0 + e^- \rightarrow \tilde e_{R,L}  \rightarrow X
$$
(see  Fig.~\ref{feyn_diag}). Here, $X$ stands for some visible
final state  arising from the $\tilde e_{R,L}$ decay, that will
be eventually revealed by the detector. 
As far as the scalar mass
is not too heavy and $\chi_1^0$ has a relevant gaugino component, 
for resonant $\chi_1^0$ energies this channel
 gives the dominant contribution to the scattering cross section,
and one can neglect $t$-channel effects.
\\
The kinematics and signature  is somehow similar to the $W^-$
resonant production  in UHE antineutrino-electron scattering \cite{glashow}
$$
\bar \nu_e + e^- \rightarrow W^-  \rightarrow X \; ,
$$
that is expected to occur at the
$Glashow$ resonant  energy $E_{\nu} = M_W^2/(2  m_e) \simeq 6.3 $ PeV.
 Since we can predict the exact characteristics of this channel
  (like its hadronic and electromagnetic showering and 
  its final visible energies),
  the corresponding  events could be used to  calibrate
the  SUSY resonance observation.

Right/left-squark ($\tilde q_R/\tilde q_L$) 
resonances can be produced through
the neutralino scattering off the quarks in the 
detector nucleons 
$$
\chi_1^0 + q \rightarrow \tilde q_{R,L}  \rightarrow X \; .
$$
Although, at the partonic level, this process has the same resonant
structure as the selectron channel,
 due to the quark momentum spread inside the nucleon, the cross-section
 resonance peak is smeared, and the $\chi_1^0$ incident-energy resonant 
 characteristics are lost.

The background coming from the UHE $\nu$ interaction with nucleons
 ($\nu + q  \rightarrow X $) by $t$-channel processes could create a major 
  continuous background  overlapping or overshadowing  the
 squark signal. In the following, after computing the relevant $\chi_1^0$ 
 event rates, we will address the issue of disentangling the SUSY
signal from  the SM resonant and continuous background given
by the competitive UHE neutrinos.

\vskip 0.2cm
Higher-energy neutralinos [with $E_{\chi}\sim m_{\tilde \nu }^2/(2 m_{\nu})$]
could interact with light relic neutrinos in the
Local Group or Super-Galactic Plane, and produce sneutrino  $\tilde
\nu$ resonances in the process \cite{Fargion:2001pu}
$$
\chi_1^0 + \nu \rightarrow \tilde \nu  \rightarrow X \; .
$$
 The following sneutrino decay  into final states containing hadrons 
 can in principle
mimic a  {\it Z-burst} event \cite{zburst}. In this paper,
we will also go through  a $\tilde \nu$-{\it burst} scenario.  

\vskip 0.4 cm
While calculating the neutralino scattering cross-sections, we will
stick to a particular model of SUSY breaking, namely the minimal
supergravity
 mediated model (mSUGRA) \cite{msugra}. The mSUGRA and its variants
are theoretically well motivated. The parameter space of mSUGRA is
characterized by five quantities: $m_0$ and $m_{1/2}$, the
universal scalar and gaugino masses, respectively; $A_0$, the
universal trilinear scalar coupling (all taken at the scale of
grand unification); $\tan \beta$, the ratio of the vacuum
expectation values of the two Higgs doublets; the sign of $\mu$,
the Higgs mixing parameter  (${\cal W} \ni \mu H_1 H_2$, where
$H_1, H_2$ are the two higgs superfields in the superpotential
$\cal W$).  While the sign of $\mu$ is arbitrary, its magnitude
is determined by requiring the radiative electroweak symmetry
breaking.

In most of the mSUGRA parameter space, the lightest neutralino is
almost a pure bino $\tilde B$, the superpartner of the $U(1)$
gauge field \footnote{In general, a neutralino is a linear
combination of the two Higgs superpartners $\tilde H_1$, $\tilde
H_2$ (higgsinos), a  $\tilde W_3$ (wino) and a $\tilde B$ (bino).}. This is
particularly important for $\chi_1^0$ being a good candidate for
cold dark matter \cite{dark_mat}. Then, the {\it
right} slepton  and  {\it right} squark  are  maximally coupled
to the lightest neutralino, and resonant $\chi_1^0 f\to \tilde f_R$ 
cross sections 
 are in general enhanced with respect to
$\chi_1^0 f\to \tilde f_L\;$ cross sections (where $f=e,q$). 

We will then
restrict our study to the right-scalar resonant production and
decay into the same initial particles $\chi_1^0 f$. 
The corresponding production rates will turn out to be 
quite straightforward to compute, the only critical parameters 
being the physical
masses of the particle involved in the process (that is the
resonance and neutralino masses, $m_{\tilde f_R}$ and $m_\chi$).
\\
In any case, one should keep in mind that the model predicts 
further new channels associated to the left scalar resonances 
(sort of \textit{twin shadows} of the right scalars) 
whose rate is in general suppressed with respect to the right scalar by
both the lower (and in general more model-dependent) relevant
decay branching fractions (see Section 3), and the heavier 
resonance masses
(since in general $m_{\tilde f_L}\gsim m_{\tilde f_R}$).

The neutralino energies that are relevant for the resonant reactions
$\chi_1^0 q\to \tilde q$, $\chi_1^0 e\to \tilde e$, $\chi_1^0 \nu\to \tilde \nu$ 
are strictly determined
by the resonance and target masses [$E_{\chi}\sim m_{\tilde
f}^2/(2 m_{target})$], that is
\bea
 E_{\chi}&\gsim& \frac{m_{\tilde q}^2}{2 m_p}\sim 10^5 \;\; {\rm GeV}  ,
 \label{rang}\\
 E_{\chi}&\sim& \frac{m_{\tilde e}^2}{2 m_e}\simeq 10^7 \div 10^9  \;\; {\rm GeV} ,
 \label{range} \\
 E_{\chi}&\sim& \frac{m_{\tilde \nu}^2}{2 m_{\nu}}\simeq 10^{14}  \;\; {\rm GeV},
\eea
respectively ($m_p,m_e,m_\nu$ being the proton, electron, relic-neutrino
masses), where we assumed $m_{\tilde q}> 300$ GeV, 
$m_{\tilde e}\simeq 100-1000$ GeV,
$m_{\tilde \nu}\simeq 200$ GeV, and $ m_{\nu}\simeq 0.1$ eV.

The plan of this paper is the following. In Section 2, we
discuss our assumptions on the  UHE-neutralino spectrum. 
 In Section~3, we present  the
cross-sections for the UHE-neutralino interaction with
electrons and quarks leading to selectron and squark resonances.
In Section~4,  we  present the event rates for the
selectron and squark resonant production by UHE $\chi_1^0$ for a
1-km$^3$ ice detector, in benchmark mSUGRA
scenarios that takes into account cosmological bounds on 
the neutralino relic
density. They will then  be compared with the number of
events expected to arise from the {\it Glashow} resonance 
$\bar\nu_e e \rightarrow W^-$,
 and from the $\nu (\bar
\nu )-$nucleon interaction via charged- and neutral-current
processes, assuming comparable UHE-neutrino and
neutralino spectra. 
In  Section~5, we  address the issue of 
disentangling the SUSY events in the detector by looking at their
peculiar energy spectrum and fluence for visible decay products.
In  Section~6, we will discuss a possible {\it $\tilde
\nu$-burst} scenario, corresponding to the resonant interaction
of  UHE $\chi_1^0$'s with light relic neutrinos in the Local Group
or SuperCluster hot dark halos. Finally,  in Section~7, we
present our conclusions.


\section{The UHE-neutralino spectrum}
In this section, we will discuss our assumptions regarding the
neutralino flux in the UHECR.  In our study, we are mainly concerned with
models of SUSY-driven decays of super-heavy particles. 
Different studies in the literature show  that the final
neutralino spectrum is quite dependent on the specific 
fragmentation model \cite{susy_oth,toldra,drees}. Here,
we will not
adopt any specific  fragmentation model, but we will follow a more
phenomenological approach.  We will try to keep our study as model 
independent as possible  by assuming for
the neutralino flux  a $-1.5$ spectral index, that should be typical of the 
fragmentation
of heavy objects. The results of a switch to a $-2$ index are also
analyzed, in order to estimate the effects of the expected uncertainty
in the predicted spectral index.
Regarding the flux normalization, we will be only guided by present
and projected experimental limits on UHE neutrino fluxes.

Neutrinos and neutralinos of comparable energies , while interacting
with matter, will produce somewhat similar signals in an earth-bound
detector. In general, neutralino-interaction rates are expected to be
smaller than the neutrino ones. This implies \textit{less
stringent} experimental bounds on neutralino fluxes with respect to
neutrino fluxes. Although there would be some room for a more abundant
incoming UHE-neutralino flux, we will assume, for the sake of
simplicity and with a conservative attitude, comparable UHE neutralino
and neutrino fluxes, that is $\phi_{\chi} \sim \phi_{\nu} $. We will
then directly turn present neutrino flux bounds into neutralino flux
limits.  Concerning the range of energies we are interested in for
selectron and squark resonances (i.e., $E_{\chi} \sim (1 - 10^3)$
PeV), AMANDA gives the most stringent experimental limit on neutrino
fluxes for energies lower than 5 PeV \cite{amanda}. In the higher
range, upper bounds on the UHE neutrino flux comes also from AGASA
\cite{agasa}, Flys Eye \cite{flys_eye},  and  RICE \cite{rice}. 

In particular, from the study of a possible excess in cascade events and
assuming relative flavor abundances $\nu_e :\nu_\mu :\nu_\tau = 1:1:1$,
AMANDA sets a limit on the total neutrino flux of
$E_{\nu}^2\;\frac{d N}{dE_{\nu}}\simeq 
8.6 ~\times 10^{-7} ~\rm GeV \;cm^{-2}\; s^{-1}\; sr^{-1},$ for
50 TeV$<E_\nu<$ 5 PeV, and  $
\simeq 1.5 ~\times 10^{-6} ~\rm GeV\; cm^{-2} \;s^{-1} \;sr^{-1},$
for 1 PeV$<E_\nu<$ 3 EeV \cite{amanda}.

For comparable interaction strength  of 
neutrinos and neutralinos with matter,   neutralino fluxes as high as
the above limits are allowed, while weaker neutralino interactions 
allow even larger spectra.

In our study, we will assume that the neutralino
flux is comparable with the flux of just one neutrino species, 
{\it i.e.} $\phi_\chi \simeq \phi_{\nu_i + \bar \nu_i}$ $(i = e,\mu,\tau)$. 
This hypothesis  is somehow conservative, 
since, in models of super-heavy particle 
decays, one can have  scenarios
where the neutralino fluxes are even larger than the sum
of all the neutrino components in the upper energy range \cite{drees}.

\vskip 0.2 cm

Regarding the neutralino flux shape (or spectral index), 
we will  model it according to the typical behavior
of present predictions  for
neutralino fluxes arising from the fragmentation of super-heavy objects
\cite{susy_oth,toldra,drees}.

 Models of UHECR fluxes arising from super-heavy $X$ particle
decays have been mainly motivated by the GZK puzzle. 
In order to have an appreciable $X$ decay rate today, one
has to tune the $X$ lifetime to be longer (but not too much
longer) than the age of the universe, 
or else {\it store} short-lived $X$-particles in topological vestiges
of early universe phase transitions. 
Then, the estimate of the $X$ lifetime and volume density is 
very model dependent.
Furthermore, the internal
mechanisms of the decay and the detailed dynamics of the first
secondaries do depend on the exact nature of the particles.

Consequently, {\it no firm prediction} on the expected flux of
neutralinos can be made. However, if there are no new mass scales
between $M_{\rm SUSY} \sim 1$~TeV and $M_X,$ the squark and
sleptons behave like their corresponding supersymmetric
partners. Then, one can infer the gross features of the $X$-particle
cascade from the {\it known} evolution of
quarks and leptons.

The upper edge of the neutralino distribution would be then
connected to the super-heavy particle mass $M_{X}$. 
In the analysis of a super-massive object
decay in \cite{drees}, the neutralino spectrum goes down with the
energy as $E^{-1.3}$, and extends up to about $E\sim M_X/2$.

At large energy fractions $x \equiv 2 E/M_X$,
the $\chi^0_1$ flux tends to be comparable or even larger
that the $\nu$ flux. 
In particular, for 
$x> (0.1-0.5)$ (depending on the
particular fragmentation treatment
\cite{susy_oth,toldra,drees}) neutralino and neutrino fluxes are
in general comparable, while at $x  < 0.1$
neutralinos are  depleted with respect to neutrinos.
Also, the total energy carried away by the
neutralinos can be comparable with that of the neutrinos.  In fact, when
the {\it initial decay} of the X particle is into SUSY particles,
the total energy of the decay carried by the neutralinos can be even 
larger than the neutrino energy (see, e.g., Tables 2.1, 2.2 and 2.3
in the last reference in \cite{drees}). 

Consequently, the neutralino flux in the energy range relevant for resonant
production of selectron and squarks [i.e. $E_{\chi} \sim 10^{6-9}$ GeV,
see eq.~(\ref{rang}),(\ref{range})], could be quite depleted when supposing 
$M_{X}\sim 10^{12-16}$ GeV in order to explain the GZK
puzzle, implying  neutralino energy fractions $x  < 0.001$.

For instance, assuming $M_{X}\simeq 2 \times 10^{12}$ GeV  to 
fit the GZK cosmic rays data (see Fig.~1 in \cite{pnath}), and
using the fragmentation model in \cite{drees}, one gets
neutralino fluxes of the order of
$E^2 _{\chi}\;\frac{d N}{dE_{\chi}} = 10^{-8}\; {\rm GeV
\;{cm^{-2}\;s^{-1}\;sr^{-1}}}$ at  
$E_{\chi}\simeq 10^{2}$ PeV (see Fig.~2 in \cite{pnath}).
This flux would be more than one order of magnitude smaller
than the value that can be extrapolated
from  present experimental upper bounds on neutrino fluxes at lower energies,
assuming, as in \cite{pnath}, a $E^{-1.3}$ spectral behavior.

On the other hand,  a lighter X-particle mass in the range $10^{8-9}$ GeV
(as considered for instance in \cite{Birkel:1998nx}) would provide a more
appropriate setting for resonant selectron and squark production, 
enhancing the neutralino flux up to a  level close to
neutrino fluxes in the $10^{7-9}$ GeV energy range. 
A $M_X$ value in the $10^{8-9}$ GeV 
 ballpark would  not be in conflict with any theoretical or
experimental consideration, although it would be less interesting
 in view of a possible  explanation of the GZK puzzle. 
 
Anyhow, we stress that at the present stage one could also think
of models the can reconcile the `low' energy range needed for
resonant selectron and squark scattering with the `large'
$M_X$ value fitting the GZK energy range.
For instance, assuming
for the heavy $X$ object a lifetime shorter than the one of the
Universe, one would presently expect a tail of redshifted secondaries.
If the $X$ decay occurs smoothly at redshift $10^{3}\gsim z \gsim 1$,
then the peak $X$ decays at largest redshift ($z\sim 10^{3}$) 
even in the case $M_{X}\sim 10^{12}$ GeV would eject its neutralino component
 at a lower energy of the order $E_\chi\simeq M_{X}/[2(1+z)] \simeq
10^{9}$ GeV today.

\vskip 0.2cm

All the previous considerations converge in our choice of a
``reference'' neutralino flux  given by 
a fragmentation-like power law
 with spectral index $-1.5$, and normalization well compatible with the 
 AMANDA bounds :
\begin{equation}
E^2 _{\chi}\;\frac{d N}{dE_{\chi}} = 10^{-7}\; {\rm GeV
\;{cm^{-2}\;s^{-1}\;sr^{-1}}} \left[
\frac{E_{\chi}}{10^{7}\;GeV} \right] ^{0.5} \; .\label{flux1}
\end{equation}
Although one could easily build up models matching the 
above ``reference'' normalization,
in the following analysis it will be straightforward to rescale
our results on expected event numbers according
to a possible more conservative (smaller) normalization factor 
in eq.~(\ref{flux1}).

 
In order to assess the effect of the uncertainty in the  spectral index
arising from  differences in modeling the super-heavy object
fragmentation, we also considered, for comparison,
a Fermi-like neutralino flux with spectral index $-2$  :
\begin{equation}
E^2 _{\chi}\;\frac{d N}{dE_{\chi}} = 10^{-7}\; {\rm GeV
\;{cm^{-2}\;s^{-1}\;sr^{-1}}}, \label{flux2}
\end{equation}
whose normalization is also compatible with experimental neutrino limits.
We than compared event numbers relative to eqs.~(\ref{flux1})
and (\ref{flux2}).

Summarizing this section, 
the details of UHE neutrinos and neutralinos fluxes  are
quite model dependent, and we are in a very preliminary understanding of
their possible spectra. Hence, we decided to use two phenomenological 
assumptions for the UHE neutralino spectrum 
differing by the spectral index, and reflecting  possible 
differences in modeling the fragmentation of a super-heavy object.
In both cases, the flux normalization
has been fixed on the basis of present
 UHE neutrino experimental bounds. 
 This hypothesis could be quite conservative, since the lower
 neutralino-nucleon cross section
 ($\sigma_{\chi N} \lsim \frac{1}{10}\sigma_{\nu N}$, see Section 4)
 would make any possible astrophysical source
 of UHE neutrinos and neutralinos more transparent to 
 (and hence more efficient in producing) the $\chi_1^0$ component.
 
 Assuming a different normalization in eqs.~(\ref{flux1})
and (\ref{flux2}) will simply imply the 
rescaling of the event numbers obtained in our final analysis.

\section{Resonant cross sections for
 UHE neutralino interactions with matter}

\begin{figure}[t]
\begin{center}
\hspace{-1in} \centerline{\hspace*{3em}
\epsfxsize=8cm\epsfysize=4cm \epsfbox{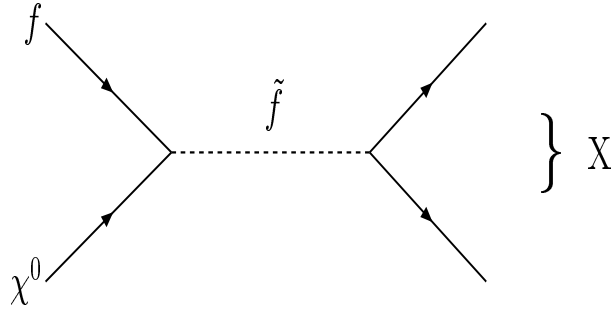} }
\end{center}
\caption{Feynman diagram of the process  $\chi_1^0 + f
\rightarrow {\tilde f} \rightarrow X$, where $f$ can be an electron, a quark
or a neutrino.}
\label{feyn_diag}
\end{figure}

UHE neutralinos with proper energy will interact with the electrons and quarks
inside the
detector, and produce selectron or squark resonances via $\chi_1^0 + (e^- , q)
\rightarrow ({\tilde e}, {\tilde q}) \rightarrow X\;$ (see the Feynman
diagram in Fig.~\ref{feyn_diag}, where $f$ is either a lepton or a quark).
The symbol $X$ stands for
a given {\it visible} final state  arising from the $\tilde f$ decay.

Near the resonance, the cross-section $\sigma$ of the process can be
written in terms of the particle masses
and the  $f \chi_1^0 $ center-of-mass (c.m.)
energy $s$ in the Breit-Wigner approximation
\begin{equation}
\sigma (s) = 2 \pi\;\frac{1}{\vert \bf k \vert ^2}\;
\frac{s\;\Gamma_{\tilde f} ^2}
{(s - m_{\tilde f}^2) + m_{\tilde f}^2 \Gamma_{\tilde f} ^2}\;
B (\tilde {f} \rightarrow f\;\chi^0_1)\;
B (\tilde {f} \rightarrow X) \;.
\label{sigma_s}
\end{equation}
In the equation above, $m_{\tilde f}$ is the slepton/squark mass,
$\Gamma_{\tilde f}$ is its total decay width, and
$B (\tilde {f} \rightarrow f\;\chi^0_1)$,
$B (\tilde {f} \rightarrow X)$ are their decay branching fractions
[$B(\tilde {f} \to\ldots)\equiv
\Gamma(\tilde {f} \to \ldots)/\Gamma_{\tilde f}$]
into the final states $f\;\chi^0_1$ and  $X$, respectively.
The modulus of the c.m. 3-momentum of the initial particles ${\bf k} $
can be expressed  as
\begin{equation}
\vert {\bf k} \vert = \frac{m_{\tilde f}^2 - m_{\chi}^2}{2 m_{\tilde f} }
\end{equation}
where $m_{\chi}$ is the neutralino mass.
Then, the c.m. energy $s$ is related to the incident neutralino energy
$E_{\chi}$ in the laboratory  by
\begin{equation}
s =  m_{\chi}^2 + 2 m_f E_{\chi},
\end{equation}
where $m_f$ is the  target  mass.
At the resonance ($s \sim m_{\tilde f}^2$), the neutralino
energy is then
\begin{equation}
 E^{peak}_{\chi} \simeq \frac{m^2_{\tilde f}- m_{\chi}^2}{2 m_f},
 \label{epeak}
\end{equation}
which sets the neutralino energy scales relevant to our study: PeVs-EeVs
for squarks and selectrons, and hundreds of ZeVs for {\it sneutrino}
bursts.

The peak cross section $\sigma^{peak}\equiv\sigma (s=m^2_{\tilde f})$
can be obtained from
 eq.~(\ref{sigma_s}) in a straightforward way :
\begin{equation}
\sigma^{peak} = \frac{8\;\pi}{m_{\tilde {f}}^2} \;
\left(\frac {m^2_{\tilde f}}{m_{\tilde {f}}^2 - m_{\chi}^2} \right)^2 \;
B (\tilde {f} \rightarrow f\;\chi^0_1)\;B (\tilde {f} \rightarrow X).
\label{sigma_peak}
\end{equation}
In the final event-rate estimate, a crucial quantity will be given
by the product $\sigma_{peak}\;\Gamma_{\tilde f}$,
the resonance peak cross section times its total decay width
(the latter setting the actual energy width of the resonance curve),
that in general
will depend on all the 5 mSUGRA
parameters.

In this analysis, we will concentrate on some important
scenarios where a relatively small parameter dependence is left
in the cross sections,
apart from the leading $m_{\tilde f}$ and $m_{\chi}$
mass dependence. In particular,
in these (quite general) scenarios, there will be little parameter
dependence in the ${\tilde f}$
decay branching fractions into both the initial $f\chi^0_1$ and final $X$
states, these branching fractions being the crucial
parameters entering $\sigma_{peak}$.
These conditions most naturally realize  in  processes
where the resonance is maximally coupled to  the initial $f\chi^0_1$
state [that is $B (\tilde {f} \rightarrow f\;\chi^0_1)\simeq 1$],
 and, at the same time,
 one considers as decay products just the same initial $f\chi^0_1 $ particles,
in the process
\begin{equation}
\chi_1^0 + f
\rightarrow {\tilde f} \rightarrow \chi_1^0 + f\;.
\label{bbuno}
\end{equation}
Then, the corresponding peak cross section
becomes 
\begin{equation}
\sigma^{peak} \to  \frac{8\;\pi}{m_{\tilde {f}}^2} \;
\left(\frac {m^2_{\tilde f}}{m_{\tilde {f}}^2 - m_{\chi}^2} \right)^2 \;.
\label{sigma_peak0}
\end{equation}
In the same scheme, also the parameter dependence of the total width
 $\Gamma_{\tilde f}$
 will be mainly restricted to phase-space mass effects.

A natural framework were to implement the above picture is given by
the {\it right selectron} and  {\it right squark} resonances
in mSUGRA.
In  mSUGRA, the  right selectron $\tilde e_R$ is in general one of
the lightest superpartner, and the lightest neutralino is mainly
a bino $\tilde B$.
For conserved R parity, the dominant $\tilde e_R$ decay channel
is  $\tilde e_R \to e\chi_1^0 $,
with a corresponding  branching fraction of almost 100 \%.
On the other hand, the left selectron $\tilde e_L$
is usually more coupled
to the second lightest neutralino $\chi_2^0$ and to the lightest
chargino $\chi_1^+$. Hence, when allowed by phase-space, the decays
$\tilde e_L \to \chi_2^0 e$ and $\tilde e_L \to \chi_1^+ e$ tend to be
dominant. Therefore, $B (\tilde {e}_L \rightarrow e\;\chi^0_1)$
is in general not very large, and
 depletes the peak $e\;\chi^0_1\to\tilde {e}_L\to X$ cross section.
\\ On the other hand, for the right selectron, according to
eq.~(\ref{sigma_peak}),
the peak cross-section is  maximized
when looking to at a $\chi_1^0  e$ final state in the channel
\begin{equation}
\chi_1^0 + e
\rightarrow {\tilde e_R} \rightarrow \chi_1^0 + e\;,
\end{equation}
 being $m_{\tilde e_R}$ and  $m_{\chi}$ the only relevant parameters
 that govern the cross section, as from eq.~(\ref{sigma_peak0}).
 
 Then, the dominant signature of the resonant interaction of a
 neutralino with an electron in the detector will be an energetic electron
 carrying away a fair fraction of the initial neutralino resonant energy.
 {\it The $\tilde e_R$ signature will be totally
 characterized by an electromagnetic
 showering, while the Glashow $W$ resonance will be mostly proceeding
 through hadronic channels.}

\vskip 0.2cm
For the {\it squark} sector, that is relevant in the neutralino interactions
with the nuclei of the detector, a similar discussion holds
regarding the  branching fractions of  the right and
left squarks into $q \chi^0_1\;$
[in particular, one has $B (\tilde {q}_R \rightarrow q\;\chi^0_1)\sim 1$,
$B (\tilde {q}_L \rightarrow q\;\chi_1^\pm)\sim 2/3$, and
$B (\tilde {q}_L \rightarrow q\;\chi_2^0)\sim 1/3$],
unless the gluino is lighter than the squark.
In the latter case, the strong interacting
decay $\tilde q_{R,L} \to q\tilde g $
is by far dominant, and the effective squark coupling to the initial
$q\tilde \chi^0_1 $ is suppressed [due to the small branching fraction
$B (\tilde {q} \rightarrow q\;\chi^0_1)\sim
\Gamma(\tilde {q} \to q\;\chi^0_1)/\Gamma(\tilde {q} \to q\, \tilde g)]$.
However, (as noted above) 
$\sigma_{peak}$ is not the only relevant quantity for
event rates, since it enters the latter through 
the product $\sigma_{peak} \Gamma_{\tilde {q}_R}$
that also involves the total squark width.
 With the onset of the
gluino decay channel, the decrease of 
$B (\tilde {q_R} \rightarrow q\;\chi^0_1)$ in eq.~(\ref{sigma_peak})
is compensated by the corresponding increase in the total
width $\Gamma_{\tilde {q}_R}$. 
Hence, when including the gluino decay $\tilde {q_R} \to q\, \tilde g$
into the {\it visible}  final states of the process
[hence keeping $B(\tilde {q}_R \rightarrow X)\sim 1$]
the final event rate will be the same as in the case of 
the lighter-than-gluino squark.

In the following we will concentrate on the resonant
production of {\it up} and {\it down} 
right squarks, $\tilde u_R$ and $\tilde d_R$,
in the processes
\begin{equation}
\chi_1^0 + u,d
\rightarrow \tilde u_R,\tilde d_R \rightarrow \chi_1^0 + u,d \;,
\end{equation}
assuming $B (\tilde u_R,\tilde d_R \rightarrow \chi_1^0 + u,d)\simeq 1$.
The dominant signature in the detector will then be given by
a strongly interacting particle initiated shower,
carrying some relevant fraction of the initial $\chi_1^0$ energy.

\section{Expected event rate at IceCube in mSUGRA benchmark scenarios}

In mSUGRA with conserved $R$ parity, the lightest neutralino is stable 
and provides an ideal 
candidate for explaining the origin of the cold dark matter (CDM).
Taking into account the  constraints on the CDM density obtained
by combining  the precise measurement of the cosmic microwave background
by the WMAP experiment \cite{wmap} and other cosmological data 
(that give $0.094 \leq \Omega_{CDM}h^2 \leq 0.129$ at 2-$\sigma$ level),
one should then require that the relic density of the lightest 
neutralino from primordial universe be compatible with $\Omega_{CDM}h^2$.
In particular, requiring that most of the CDM is made up of lightest 
neutralinos restricts quite a lot the allowed region of the
mSUGRA $(m_0,m_{1/2})$ parameter plane (see, e.g., \cite{Olive:2005qz},
and references therein).

In \cite{Battaglia:2003ab}, a set of 13 mSUGRA benchmark scenarios
(characterized by a given set of values for the mSUGRA parameters $m_0$,
$m_{1/2}$, $A_0$, at the GUT scale, and by the values of $\tan\beta$ and
$sign\,\mu$)
have been chosen, respecting the following experimental constraints:
non-observation of SUSY partners at colliders (especially at LEP)
\cite{emass,tevatron,lep_limits}; 
bounds on deviation from  the
SM $b\to s\gamma$ decay rate; 
agreement of the lightest-neutralino relic density
with $0.094 \leq \Omega_{CDM}h^2 \leq 0.129$.

Of the 13 benchmark points in \cite{Battaglia:2003ab}, 
5 points (B', C', G', I', L') are in the `bulk'
region (i.e. at moderate $m_0$, $m_{1/2}$ values),
4 points (A', D', H', J') are along the co-annihilation `tail'
(at moderate $m_0$ and larger $m_{1/2}$, where the neutralino-stau 
co-annihilation channel considerably affects the neutralino relic density),
2 points (K', M') are along rapid-annihilation `funnel' 
(where both $m_0$ and $m_{1/2}$  can grow large with large $\tan\beta$),
and 2 points (E', F') are in the `focus-point' region, at very large $m_0$.

In the present analysis, the two crucial parameters ruling the signal rates
are the resonance scalar mass $m_{\tilde f}$ 
and the splitting in the square of the
scalar and neutralino masses, $m^2_{\tilde f}- m_{\chi}^2$. 
Then, we found that scenarios like
K', M' (`funnel') and E', F' (`focus points'), where the scalar mass 
parameter is quite large ($m_0 \geq 1$ TeV $\Rightarrow m_{\tilde f} > 1$ TeV), 
give rise to drastically reduced  signals (less than 0.01 events/year
in Icecube, see below).
On the other hand, for such large scalar masses, $t$-channel  diagrams
in the $\chi_1^0 f$ scattering can become dominant over the resonant
$s$-channel,
in case neutralino has a non-negligible higgsino component (cf. `focus-point'
scenarios). Then, the  approximation
of restricting the cross section evaluation to the resonant channel
(on which the present analysis in based) could fail.
Hence, in our study, we will consider only the `bulk' and 
co-annihilation benchmarks, leaving the analysis of the 
`funnel' and `focus-point' 
scenarios to a more complete treatment.
\begin{table}[t]
\centering
\renewcommand{\arraystretch}{0.90}
{~}\\
\begin{tabular}{|c|r|r|r|r|r|r|r|r|r|}
\hline
Model          & A  &  B &  C &  D &  G &  H &  I &  J &  L   \\ 
\hline \hline
$m_{1/2}\;\rm (GeV)$  & 600 & 280 & 400 & 525 &   375 & 935 & 350 & 750 &450  \\
$m_0 \;\rm (GeV)$ & 107 &  57 &  77 & 101 &  110 & 233 & 180 & 298 &303  \\
$\tan{\beta}$  & 5   &  10 &  10 & 10  &  20  & 20  & 35  & 35  &  47    \\
sign($\mu$)    & $+$ & $+$ & $+$ & $-$ &  $+$ & $+$ & $+$ & $+$ & $+$   \\ 
\hline
$\Omega h^2$    & 0.128 & 0.123 & 0.122 & 0.116 &  0.110 & 0.123 & 0.117 & 0.119 
&0.113   \\ 
\hline \hline
$m_{\chi} \;\rm (GeV)$  & 243 &  107 & 158 & 212 & 148 & 388 & 138 & 309 & 181 \\
$m_{\tilde e_R}\;\rm (GeV)$ & 254 & 128 & 175 & 226 & 184 & 423 & 227 & 412 & 349  \\
$\Gamma_{\tilde e_R}\;\rm (MeV)$  & 9.13 & 57.2 & 29.7 & 16.3 & 103 & 53.4 
& 448 & 396 & 931 \\
$\sigma^{peak}_{\tilde e_R}\;$ ($\mu$b)  & 21.1 & 6.58 & 9.34 & 13.3 
& 2.57 & 2.17 & 0.477 & 0.301 & 0.150 \\
\hline \hline
{${\cal N}_{\tilde e_R}$} 
& 28  & 32& 26 & 23 & 8.5 & 2.3 &1.9 & 0.55 & 0.42   \\ 
{(${\rm km}^{-3}\;{\rm yr}^{-1})$}
& 37& 45& 34& 30& 7.7
& 1.4& 1.1& 0.20& 0.14 \\ 
\hline \hline
$m_{\tilde q_R}\;\rm (GeV)$ &1194 & 596 & 825 &1057 & 781 &1798 & 748 &1487 & 961  \\
$\Gamma_{\tilde q_R}\;\rm (GeV)$  &2.45 & 1.23 & 1.70 & 2.17 & 1.61 & 3.65 
& 1.55 &3.04 & 1.99  \\
$\sigma^{peak}_{\tilde q_R}\;\rm (nb)$   &7.43 & 28.6 & 15.3 &9.46 & 17.1 & 3.32 & 18.5 
&4.81 & 11.3  \\
\hline \hline
{${\cal N}_{\tilde q_R}$}  & 0.35 &3.0 &1.1  &0.52  &1.3  &0.08 &1.5 &0.17 
&0.70\\
{(${\rm km}^{-3}\;{\rm yr}^{-1})$}
& 0.22& 2.6& 0.89& 0.36 & 1.1& 0.04& 1.2& 0.09 & 0.51 
\\
\hline
\end{tabular}
\caption{\label{events} 
Definition of different mSUGRA scenarios and corresponding event-number
expectations ${\cal N}_{\tilde e_R,\tilde q_R}$, for resonant neutralino 
scattering in ice.
In all scenarios, we assume $A_0 = 0$.
The resonance decay widths and peak cross sections for the right selectron 
and right squark are also shown.  The upper of the two
entries in the ${\cal N}_{\tilde e_R,\tilde q_R}$ raws 
corresponds to the event number
calculated by the neutralino flux in
eq.~(\ref{flux1}) ($\beta_0=1.5$), while the lower entry refers to
eq.~(\ref{flux2}) ($\beta_0=2$).  When computing the squark
event rates, an incident neutralino energy threshold of 1 PeV is assumed.
The relic DM density $\Omega h^2$ corresponding to each scenario is 
also shown.}
\end{table}

The 9 benchmark points we include in our analysis  are reported in
Table~\ref{events}. They correspond either to the `bulk' region
(B, C, G, I, L) or to the co-annihilation region (A, D, H, J).
In all  scenarios, $A_0 = 0$.

Starting from the values of $m_0$,
$m_{1/2}$, $A_0$ at the GUT scale, and $\tan\beta$,
$sign\,\mu$, all the physical masses and couplings of the spectrum of 
SUSY partners can be calculated
at the electroweak scale by solving a set of renormalization group
equations.
There are many public spectrum calculators doing this task,
including second-order radiative corrections (see, i.e., \cite{comparison}).
The differences in the corresponding results is considered a good 
estimate of the present
theoretical uncertainty in the computation of SUSY spectra.
\\
Then, the high precision of the WMAP  $\Omega_{CDM}h^2$ determination 
(that slimmed considerably the allowed mSUGRA parameter regions) gives rise,
 through the prediction of the neutralino relic density versus
 mSUGRA parameters,
 to some sensitivity of the allowed mSUGRA regions
to the choice of the spectrum calculator in the analysis. 
The differences in results is considered a good estimate of the present
theoretical uncertainty of this kind of analysis.
\\
The impact of the sensitivity to the details of the
mSUGRA spectrum has been analyzed in \cite{comparison}
through the code micrOMEGAs 1.3 \cite{Belanger:2004yn} that can
evaluate the neutralino relic density assuming different codes for the
mSUGRA spectrum. One finds 
that mass differences of about 1\% in the mSUGRA
spectrum can imply sometimes  a 10\% variation (and even larger
for high $\tan\beta$ and $m_0$) in the corresponding 
neutralino relic density.

The above discussion motivates the choice of our benchmark points
in Table~\ref{events}.
In our analysis, we used Isajet 7.71 \cite{Paige:2003mg}
to compute the mSUGRA spectrum\footnote{In Isajet7.71, we set
$m_b(m_b)^{\bar{MS}}=4.214$ GeV, $\alpha_s(M_Z)^{\bar{MS}}=0.1172$
and the top pole mass at $m_t=175$ GeV.},
and micrOMEGAs 1.3 to compute the corresponding neutralino relic density
(as implemented in the last reference in 
\begin{figure}[H]
\vspace*{0.5truecm}
\centerline{
\mbox{\epsfxsize=14.5 truecm\epsfysize=14.5 truecm\epsffile{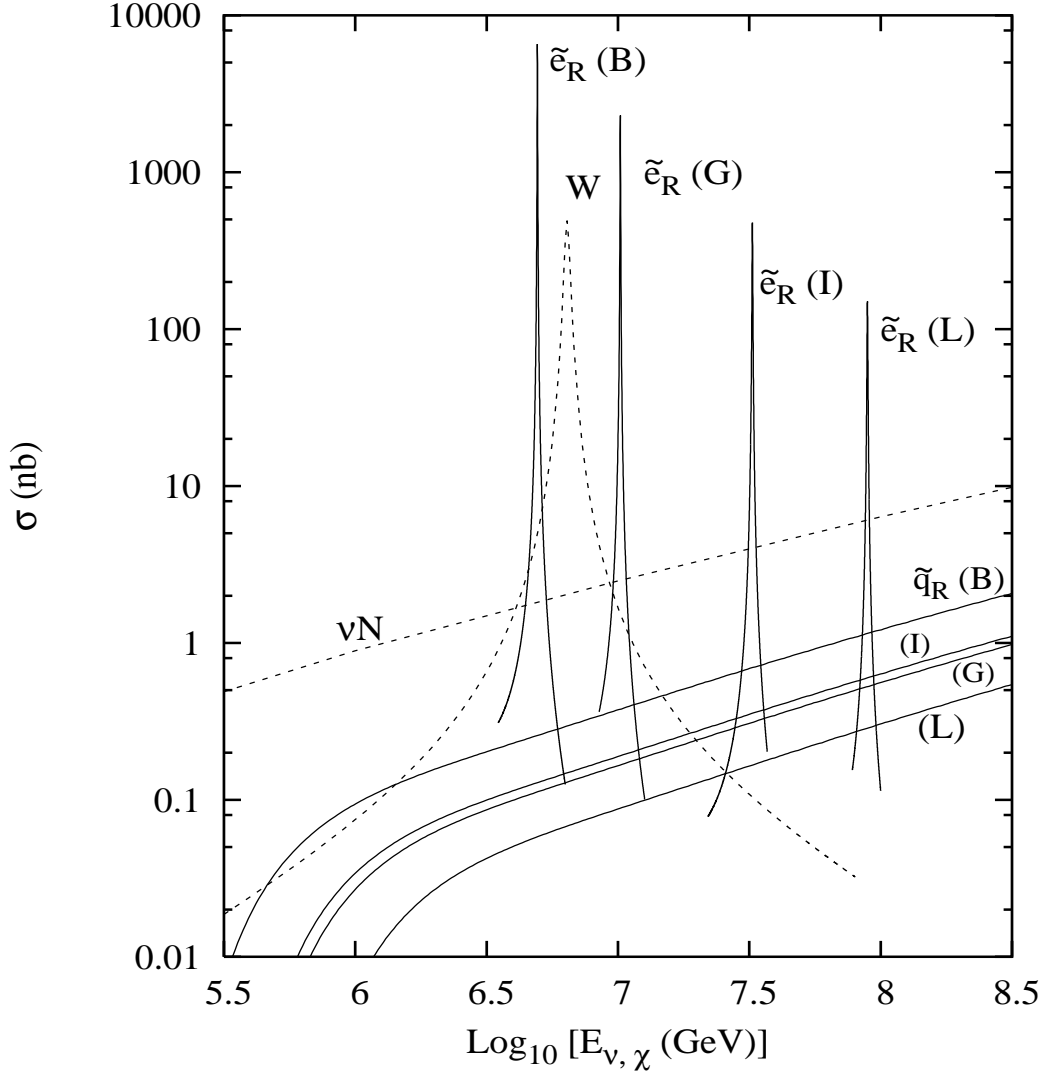}}
}
\caption{\label{reson}
 Cross sections (in nb) for 
the resonant right-selectron/right-squark production
by UHE neutralino scattering
on electrons/nucleons, in the four mSUGRA scenarios B, G, I, L defined
in Table~\ref{events}. For comparison, the dashed lines show the scattering
cross sections of UHE anti-neutrinos on electrons (when marked by $W$)
and  nucleons (when marked by $\nu N$).
}
\end{figure}
\noindent \cite{comparison}).
The small differences in the definition of our benchmarks A, B, C, $\ldots$ 
in Table~\ref{events}
with respect to the corresponding post-WMAP benchmarks
A', B', C', $\ldots$ defined in Table 2 of \cite{Battaglia:2003ab} 
(which are based 
on ISASUGRA 7.67) are due to the small adjustments in the choice of
$m_{1/2}$ and $m_0$ needed 
to pass both the relic-density and LEP constraints by assuming 
a different code. 

In Table~\ref{events}, one can read the neutralino relic density, 
$\Omega h^2$, corresponding to each scenario, that is well compatible with
the assumption that the lightest neutralino be the main CDM component.
For each benchmark, we show the lightest-neutralino, right-selectron
and right-squark masses (masses for
squark of different flavors $u,d,s,c$ are assumed to be degenerate
and equal to $m_{\tilde q_R}$). In all scenarios, one finds
$m_{\tilde e_R}< 0.5$ TeV, and 
$m_{\tilde q_R}<m_{\tilde g}$,
so that $B (\tilde {q}_R \rightarrow q\;\chi^0_1)\sim 1$.
Total right-selectron and right-squark widths, $\Gamma_{\tilde e_R}$
and $\Gamma_{\tilde q_R}$ are also shown
in Table~\ref{events}, along with the respective peak cross sections,
$\sigma^{peak}_{\tilde e_R}$ and $\sigma^{peak}_{\tilde q_R}$, as
defined in eq.~(\ref{sigma_peak}) with $B (\tilde {f} \rightarrow X)\simeq
B (\tilde {f} \rightarrow f\;\chi^0_1)$.

In Fig.~\ref{reson}, 
we show  (for the B, G, I, L scenarios)
the cross sections for the
resonant production of a right selectron and a right squark
from the UHE neutralinos scattering on electrons and nucleons, 
respectively, versus
the incident neutralino energy.

For comparison, we also present the resonant neutrino cross section for
$\bar \nu_e e^-\rightarrow W^- \rightarrow  all$, and
the (non-resonant) neutrino-nucleon cross section.
Indeed, a Breit-Wigner $W$ peak cross section is expected
at the incident neutrino energy ($E_\nu$)
\begin{equation}
 E^{peak}_\nu = \frac{M^2_{W}- m_\nu^2}{2 m_e} \simeq \frac{M^2_{W}}{2 m_e}
 \sim ~6.3 \times 10^6 \;\rm GeV \;,
\end{equation}
although present observation of UHE energy neutrino events does not reach
such energies yet.

Cross sections for the resonant squark production do not show the
Breit-Wigner structure.  Because of
the continuous quark momentum distribution  inside a nucleon $N$,
any value of the incident neutralino energy
larger than a threshold ($E_\chi> \frac{m^2_{\tilde q}- m_{\chi}^2}{2 m_N}$,
where $m_N$ is the nucleon mass)
can produce a squark at resonance.
In this case, the cross section evaluation  involves a
convolution of the partonic cross section in  eq.~(\ref{sigma_s})
(with $f \equiv u,d$ and {\it sea} quarks) with a parton distribution function
$f_q(x,Q^2)\;$, where $x$ is the momentum fraction of the proton carried by
the parton, and $Q$ is the energy scale at which we evaluate this
distribution. In our computation, we used the CTEQ4L set of 
structure functions 
in \cite{cteq}.

The $\tilde u_R$ and $\tilde d_R$ decay widths are small
compared to their masses, and one can substitute the Breit-Wigner
propagator in the partonic cross section by a $\delta$-function.
Then,   the squark-resonance cross section
in the $\chi^0_1$-nucleon scattering becomes:
\begin{equation}
\sigma_{\tilde {q}_R}(E_{\chi}) =\pi \;\sum_q \;
\frac{\sigma^{peak}_{\tilde {q}_R}\;\Gamma_{\tilde {q}_R}}{m_{\tilde {q}_R}}
\;\;\; x_q\;\;f_q (x_q, m_{\tilde {q}_R} ^2)
\label{sig_part}
\end{equation}
\noindent
where, $x_q = {m_{\tilde {q}_R}^2}/({m_{\chi^0_1}^2 + 2\;E_{\chi}
m_N})$. The summation in
eq.~(\ref{sig_part}) is over the $u$- and $d$-type quarks 
(thus including also $s$ and $c$ quarks) in an isoscalar
target.  In our scenarios, one has
$m_{\tilde {u}_R} \simeq m_{\tilde {d}_R}\simeq 
m_{\tilde {c}_R} \simeq m_{\tilde {s}_R}$
and $\Gamma_{\tilde {u}_R}\simeq \Gamma_{\tilde {c}_R}
\simeq 4\;\Gamma_{\tilde {d}_R}\simeq 4\;\Gamma_{\tilde {s}_R}\;$.

The actual
event rates ${\cal N}_{\tilde f}$ can be calculated from the
cross section by convoluting
it with the appropriate neutralino flux, and  multiplying the result by the
number of the target electrons/nucleons ${\cal N}_T$ in a given
 detector volume, and by the exposure time $\Delta t$, as follows
\begin{equation}
{\cal N}_{\tilde f} = {\cal N}_T\;\Delta t\;
\int \sigma_{\tilde f}(E_{\chi}) \;\frac{d N}{d E_{\chi}} \;dE_{\chi}\;.
\end{equation}
For selectrons and squarks
$ \sigma_{\tilde f}(E_{\chi})$  will be given by
eq.~(\ref{sigma_s}) and eq.~(\ref{sig_part}), respectively. We assumed the
two different neutralino
fluxes  in eq.~(\ref{flux1}) and eq.~(\ref{flux2}). ${\cal
N}_T$ is given by the product of the detector mass (in g units)
and Avogadro number ($N_A$) for nucleons (times 11/18 for electrons).

The selectron width being small compared to its mass (see Table~\ref{events}),
we approximated the
Breit-Wigner propagator in eq.~(\ref{sigma_s}) by a
$\delta$-function. The corresponding expected event number is then
\begin{equation}
{\cal N}_{\tilde {e}_R} =  \alpha_{\tilde e_R} \;\sigma^{peak}_{\tilde e_R}\;
\Gamma_{\tilde {e}_R} \; \left(\frac{m_{\tilde {e}_R}}{2\; m_e}\;\right)
\left({E_{\chi}^{peak}}\right)^{-\beta_0} \;,
\label{evts_er}
\end{equation}
where the factor $\alpha_{\tilde e_R}$ (depending on the mass of
the detector, exposure time and nature of the neutralino flux)
includes the product of $\Delta t$, ${\cal N}_T$, and the
normalization of the fluxes defined in eq.~(\ref{flux1}) and
eq.~(\ref{flux2}). The exponent $\beta_0$ sets the shape of the
neutralino flux, and is equal to $1.5$ and $2$, respectively, for
the two cases, while $E_{\chi}^{peak}$, expressed in GeV, is given
by eq.~(\ref{epeak}) where $f\to e$.

For an ice detector of 1 km$^3$ and one year of exposure, one has
$\alpha_{\tilde e_R}=7.24 \times 10^3$ $[2.29 \times 10^7]$
nb$^{-1}$ GeV$^{\beta_0-1}$  for the flux in eq.~(\ref{flux1})
[eq.~(\ref{flux2})]. The large difference in the $\alpha$ factors
is an artifact due to the exponent $\beta_0$ in the formula above.

\begin{table}[H]
\begin{center}
\begin{tabular}{|c|c|c|}
\hline
\multicolumn{3}{|c|}{Number of events from UHE $\nu$, $\bar \nu$ interaction} \\
\hline 
$W^- \rightarrow e \bar \nu_e$&$W^- \rightarrow \rm hadrons$&
$\sum_i (\nu_i + \bar \nu_i)_{CC + NC}$ \\
\hline
2 & 12 & 55 \\
2.5 & 15 & 56\\
\hline
\end{tabular}
\end{center}
\caption{Number of events from UHE neutrino/antineutrino CC and NC interaction 
in IceCube per year. In each entry, the upper number refers to
the  flux in eq.~(\ref{flux1}) ($\beta_0=1.5$) for
 the 
 $\nu_i+ \bar \nu_i\;$ spectrum, while the lower number
 assumes the flux in  eq.~(\ref{flux2}) ($\beta_0=2$). 
An incident-neutrino energy
threshold of 1 PeV is used.}
\label{sm}
\end{table}

For the $\tilde q_R$-resonances,
due to the energy distribution of the partons in a nucleon,
the number of events
will get contributions from a continuous
range of incident neutralino energies
\begin{equation}
{\cal N}_{\tilde {q}_R} = \alpha_{\tilde q}\;\int_{E_{\chi}^{min}} \;
\sigma_{\tilde {q}_R}(E_{\chi})\;\;E_{\chi}^{-\beta_0} \;\;dE_{\chi} \;,
\label{evts_qr}
\end{equation}
where, for a 1-km$^3$ ice detector and one year of exposure,
 $\alpha_{\tilde q_R}=
3.77 \times 10^3$ $[1.19 \times 10^7]$ nb$^{-1}$ GeV$^{\beta_0-1}$ for the
flux in eq.~(\ref{flux1}) [eq.~(\ref{flux2})], and
$\sigma_{\tilde {q}_R}(E_{\chi})$ is given by eq.~(\ref{sig_part}).

One can note that
the peak cross section in eq.~(\ref{sigma_peak0})
is enhanced for small mass differences
between the selectron/squark and the neutralino.
However,  in the product
$\sigma^{peak}_{\tilde f}\; \Gamma_{\tilde f}$ that enters the
event numbers in eq.~(\ref{evts_er}), and in eq.~(\ref{evts_qr}) through
eq.~(\ref{sig_part}),
the $1/(m_{\tilde f}^2 - m_{\chi}^2)^2$ behavior in
$\sigma^{peak}_{\tilde f}$ is
compensated by the resonance-width,
that goes as  $(m_{\tilde f}^2 - m_{\chi}^2)^2$.
On the other hand, a residual enhancement associated to the
resonance-neutralino mass
degeneracy arises from the factor $\left({E_{\chi}^{peak}}\right)^{-\beta_0}$
in eq.~(\ref{evts_er})
(cf. eq.~(\ref{epeak})), and from the lower limit 
$E_{\chi}^{min}= (m^2_{\tilde q}- m_{\chi}^2)/(2 m_N)$
of the integral in eq.~(\ref{evts_qr}).
Hence, selectrons/squarks and neutralinos that are closer in mass 
tend to increase
the expected event number.

In Table~\ref{events}, we present the number of events (${\cal
N}_{\tilde e_R ,\;\tilde q_R}$), per km$^3$ per year, expected in an
ice detector for the different mSUGRA benchmarks. 
The number ${\cal N}_{\tilde q_R}$ was always
worked out by assuming as lower limit in the integration in
eq.~(\ref{evts_qr}) the value $E_{\chi}^{min}\simeq 10^6$ GeV in order
to cut off the atmospheric neutrino pollution.  In Table~\ref{events},
the raws corresponding to the number of events have two entries.
The upper one corresponds to the neutralino flux defined in
eq.~(\ref{flux1}) ($\beta_0=1.5$), while the lower one 
assumes the flux in eq.~(\ref{flux2}) ($\beta_0=2$).

For comparison, the expected event rates for the resonant $W$
production from $\bar \nu_e e$ scattering, in the hadronic and leptonic
channels, are presented in Table~\ref{sm}. Since the fluxes
in eqs.~(\ref{flux1}) and (\ref{flux2})
are relative to one flavor of $\nu + \bar \nu$, 
assuming a complete neutrino flavor 
mixing due to non-zero neutrino masses, 
in Table~\ref{sm}
we use a $\bar
\nu_e$ flux that is half of those defined in eq.(\ref{flux1}) and
eq.(\ref{flux2}). We  also show total $\nu$-N and $\bar \nu$-N
charged-current (CC) plus neutral-current (NC) event rates.

In scenarios A, B, C, D, where one has both a light selectron
($m_{\tilde e_R}\lsim 250$ GeV) and  quite degenerate $m_{\tilde e_R}/m_\chi$
values, one obtains in IceCube a few tens of events per year in 
the electronic channel.
Note that the selectron/neutralino degeneracy  favors
${\cal N}_{\tilde e_R}$ in the harder-spectrum case with $\beta_0=1.5$.
In scenarios  H, I, J, L the larger splitting in 
$m_{\tilde e_R}$ and $m_\chi$, and the quite heavy selectron 
deplete the event rates by more then one order of magnitude (0.2-2 events per
year).
Note that in this case the $\beta_0=2$ spectrum gives higher
${\cal N}_{\tilde e_R}$.
The benchmark G shows an intermediate case, where the selectron is quite light,
while  $m_{\tilde e_R}$ and $m_\chi$ are not much degenerate, giving rise to
about 8 events per year.

Concerning the hadronic channel, due to the large squark masses
($m_{\tilde q_R}\gsim 600$ GeV) and their large splitting with $m_\chi$,
the signal is modest in all benchmarks, reaching the level of 1-3 events 
per year only in the B, C, G, I points.

With the assumed neutralino fluxes, one could expect a clear SUSY 
signal in IceCube in the
electronic channel for most of the analyzed benchmarks.
Only scenarios J and L foresee less than 1 event per year in both
the electronic and hadronic channel.
We recall that the
normalization of the fluxes in eqs.~(\ref{flux1}) and (\ref{flux2})
is quite conservative. 
Due to the lower neutralino-nucleon cross section
$\sigma_{\chi N} \lsim \frac{1}{10}\sigma_{\nu N}$ 
(cf. Fig.~\ref{reson}), one is presently allowed to
assume flux normalizations that are an order of magnitude larger than the ones
adopted here without violating any experimental constraint. 
This would make the  event statistics even more promising.

\section{Disentangling the neutralino signal by neutrino calibration}
We will now briefly address the problem of the effective {\it
visible-energy } spectrum coming from the two body decays
$\tilde f_R\to f \chi_1^0$, that is of course a crucial
characteristic in the detection of a resonance signal.

\begin{figure}[H]
\vspace*{1truecm} 
\centerline{
\mbox{\epsfxsize=14.truecm\epsfysize=14.truecm\epsffile{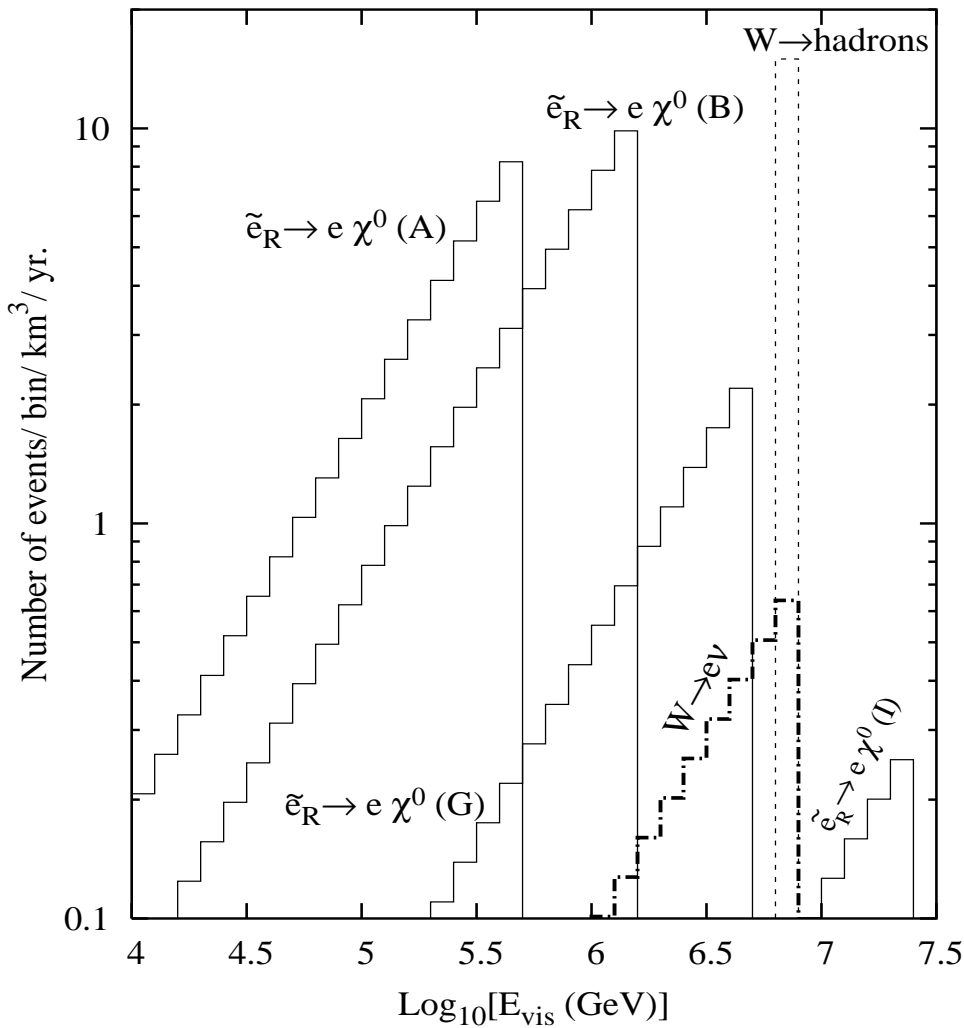}}}
\caption{\label{spectrum} Visible energy spectra of the resonance
decay products [assuming the flux
in eq.~(\ref{flux1}) ($\beta_0=1.5$)]. 
For the resonant $W$ production and 
decay, the dash-dotted (dashed) histogram represents 
the electron (hadronic) energy
spectrum in the lab. The four solid-line distributions  refer to the 
electron energy spectra arising from $\tilde e_R
\rightarrow e \chi^0$ in the mSUGRA scenarios A, B, G, I defined in
Table~\ref{events}.}
\end{figure}

In the resonant $\tilde e_R$  decay into an electron plus an invisible
neutralino $\chi_1^0$, only the electron energy contributes to
the visible energy. In Fig.~\ref{spectrum}, we plot the visible-energy
spectrum 
$\Delta z\frac{dN}{d z}\;$ (where $z=\log_{10} E_{vis}$, and $\Delta z=0.1$
assumes a 10\% experimental resolution)
corresponding to the resonant
$\tilde e_R$ production and decay into $e \chi_1^0$,
for benchmarks A, B, G, I. 
The
positions of the sharp end-points of the spectra are determined by the
resonance  and neutralino masses.
For comparison,
we also show the expected visible-energy spectrum for a resonant $W$ decaying
into either $e\nu$ or hadrons.  When the $W$ decays hadronically, the
total energy of the initial incident neutrino gives rise to
visible energy in the final state. In this case, the
$\Delta z\frac{dN}{d z}\;$
spectrum has a resonant structure
corresponding to the initial neutrino energy $E_\nu\simeq m_W^2/(2
m_e)\simeq 6.3 \times 10^6$ GeV. When the $W$
decays leptonically, only a fraction of the incident neutrino energy
goes into visible energy, whose distribution is characteristic of a
two-body decay, just as in  $\tilde e_R\to e \chi_1^0$.  
The different positions of the spectrum end-points could then provide a
handle to disentangle the visible energy  of  the $\tilde e_R$ decay
from the one of the $W$ decay.

The hadronic $W^-$ decay
defines a  sharply clustered group of events [about 12 for 
the $\beta_0=1.5$ spectrum, and 15 for the $\beta_0=2$ power law, per year,
cf. Table~\ref{sm}], 
whose
intensity may be correlated with the two electromagnetic channels
arising from  $W^- \rightarrow e^- + \bar\nu_e $ [2  and 2.5 events] and $W^-
\rightarrow \tau^- + \bar\nu_{\tau} $ [2 and 2.5 events]. The 
 channel $W^- \rightarrow \mu^- + \bar\nu_{\mu} $ [2 and 2.5 events] 
is delivering just a single PeV muon track, that is quite difficult to 
recognize. The electron and tau electromagnetic energy spectra 
 grow linearly with energy up to  $E =
6.3$ PeV, and their average energy is about
half  of the hadronic-channel energy. 
The distribution is  spread according to
 a continuous power law, as shown in Fig.~\ref{spectrum}. The
$\tau$ birth and decay will be in general source of a characteristic
{\it double-bang} event \cite{doub_bang}. 

In conclusion, the
appearance of a sharp edge at some energy different  from the 
expected $W^-$
hadronic, electronic, and tau bumps could point to some new physics
corresponding to the resonant $\tilde e_R$ production from UHECR
neutralinos. Such reconstruction seems possible 
for a neutralino flux that is comparable to (and, in favorable scenarios,
even lower than)
the  present lower bounds on neutrino fluxes.

The mass ranges considered are of great astrophysical interest, and
within the reach of the CERN LHC. A near-future underground
$\nu$-detector could then cross-check a possible SUSY discovery at
future colliders.

\section{The role of the sneutrino burst}

The resonant production of $Z$-bosons via UHE $\nu$ scattering
off massive relic neutrinos has been the backbone of the so
called {\it Z-burst} mechanism  \cite{zburst} that could explain
the  GZK puzzle. 
UHE neutrinos in {\it Z-burst} are the messengers able to overcome the GZK
cut-off, by offering a link between UHECR and most distant cosmic sources.
About 70\% of the $Z-$bosons  decays
hadronically, thus producing protons, neutrons and anti-nucleons
as well as  pions (and consequent photons from  neutral-pions
decays). The $Z$-decay products could have quite naturally
energies beyond the GZK scale for appropriate relic-neutrino
masses, since $E_{\nu} \simeq M_Z^2/(2m_\nu) \simeq 4 \cdot
10^{22}\left(\frac{0.1 eV}{m_\nu}\right)eV$, which corresponds to
a UHECR proton energy of
$E_{p} \simeq \frac{1}{20}\cdot M_Z^2/(2m_\nu)$. If  relic-neutrino 
halos are present at distances less than about 50 Mpc
\footnote{The distances of the center of dense clusters from the
earth are estimated to be 16 Mpc for Virgo, while  our
Super-Galactic Plane diameter is about 50 Mpc \cite{cluster}.},
then these protons or anti-protons  (or photons) could arrive to
the earth atmosphere without much energy loss. Consequently,
air-showers produced by these extremely high-energy nucleons could
give rise to the {\it super} GZK events, as detected by the AGASA
experiment \cite{agasa}. Crucial ingredients of this picture
are of course sufficiently large
 UHE-neutrino fluxes and relic-neutrino densities in the
Local Group or Super Galactic Plane. Then, 
the {\it Z-burst} model has a good potential
for explaining observed correlation of UHECR
arrivals from BL-Lac sources at distances well above GZK, and
electron pair losses \cite{Gorbunov02}.

In this section, we will consider
the resonant sneutrino ($\tilde \nu$) production from the UHE
neutralinos scattering off relic  neutrinos,
and its subsequent decay into visible final states
\begin{equation}
\chi_1^0  \nu \to \tilde \nu \to X_{vis} \;,
\label{sneut}
\end{equation}
where $X_{vis}$ stands for the all  final states with
at least one visible particle.
In particular, we will discuss whether this process can have a role
similar to the {\it Z-burst} in the
observed  component of the cosmic-ray spectrum
beyond the GZK energy.

The channel in eq.~(\ref{sneut}) differs from the 
previously analyzed resonant selectron and squark
production under various aspects.
First, due to the lightness of the target $\nu$ mass,  it requires
much larger $\chi_1^0$ energies according to eq.~(\ref{epeak}).
Second, optimizing cross sections by considering scenarios where the
sneutrino is maximally coupled to the initial state, and,
at the same time, it decays into the
initial particles $\tilde \nu\to\chi_1^0  \nu $
is not viable, since
this would give rise to invisible final states.
On the other hand, the
sneutrino can
decay into heavier neutralinos $\chi_i^0$ ($i=2,3,4$) and charginos
$\chi_j^\pm$ ($j=1,2$). The subsequent neutralino and chargino decays into
 hadrons and leptons might produce
highly energetic protons, pions and electrons that would be
able to initiate the
super GZK air showers.
 Looking back at eq.~(\ref{sigma_s}) for the cross section,
 one is  interested in scenarios where neither
$B(\tilde \nu\to\chi_1^0 \nu)$ nor $B(\tilde \nu\to X_{vis})$
are small.
The latter are complementary quantities, since
one can assume quite accurately
$B(\tilde \nu\to X_{vis}) \simeq 1-B(\tilde \nu\to\chi_1^0 \nu)$.
Hence, the sneutrino production cross section is automatically
penalized.
Furthermore, the rates will be in general quite model dependent.
Our results will not be simply determined by the particle mass spectrum
(as in the right selectrons/squarks production),
but  will  depend also on  couplings (and on the $\tan\beta$ and $\mu$
parameters) in a nontrivial way.
Finally, since a visible sneutrino channel will arise from a
decay chain,
the energy of the visible decay products can in principle be quite depleted.
 In the following, we will also
 discuss the sneutrino energy flow in the
visible decay products.

The sneutrino branching fraction into visible final states
$B(\tilde \nu\to X_{vis})$ is in general made up of different components
\begin{equation}
B(\tilde \nu \rightarrow X_{vis}) \equiv
\sum_{j=1}^2 B(\tilde \nu \rightarrow l \chi^\pm_j)
+ \sum_{i= 2} ^4 B(\tilde \nu \rightarrow \nu \chi^0_i)
  \;B(\chi^0_i \rightarrow f \bar f \chi^0_1 )\sim
  1-B(\tilde \nu\to\chi_1^0 \nu)
\end{equation}
where $f \ne \nu$.

The $\tilde \nu$ decays into the second neutralino $\chi_2^0$ and into
the lightest chargino $\chi_1^\pm$, when allowed by phase space, are in general
dominant.
 The subsequent decays  $\chi_1^\pm \to q
\bar q' \chi_1^0$ (with a branching ratio of about 66 \% ) and
$\chi_1^\pm \to l \nu_l
\chi_1^0$ (with a branching ratio of about 33 \% ) are the sources of hadrons
and charged leptons, respectively, 
that may trigger the highest-energy air showers.
The second neutralino  can have a substantial decay rate
into the invisible channel $\chi_2^0 \rightarrow \nu \bar \nu
\chi_1^0$. Otherwise, a $\chi_2^0$ decay  produces
quarks  and leptons via $\chi_2^0 \rightarrow q \bar q \chi_1^0, \; l
\bar l \chi_1^0$.  The $\chi_2^0 \rightarrow e \tilde e_R$
decay may also be  kinematically allowed.
In the latter case, no hadron would arise  from the decay chain.

The presence of a heavy invisible particle ($\chi_1^0$) among the
$\tilde \nu$ decay products makes the visible part of the spectrum
quite different from the {\it Z-burst} case. The results are in
general  model dependent and less promising, and the incoming UHE $\chi^0$
energy should be at higher energies. 
The decay kinematics of a {\it
Z-burst} event is quite simple. The hadrons resulting from the $Z
\rightarrow q \bar q$ decay carry away all the energy of the incident
UHE neutrino (i.e., the $Z$ resonance energy). On the other hand, the
average energy of the visible sneutrino decay products is lower
than the sneutrino mass, since hadrons can only appear either from a
three-body decay of $\chi_2^0,
\chi_1^\pm$, or from a two-body decay chain [$\chi_2^0
(\chi_1^\pm) \rightarrow Z (W) \chi_1^0$, with $Z(W) \rightarrow
\rm hadrons$] following the initial two-body $\tilde \nu$ decay.
Sneutrino masses, anyway, can be quite heavier than the $Z$ mass,
and this could compensate partly the depletion of the average energy
 of the sneutrino decay products, that in any case would
have a less {\it resonant} structure than in the {\it Z-burst} case.
\begin{table}[t]
\centering
{~}\\
\begin{tabular}{|c|r|r|}
\hline
Model          & I  &  L   \\ 
\hline \hline
$m_{1/2}\;\rm (GeV)$ & 350 &450  \\
$m_0 \;\rm (GeV)$ & 180 &303  \\
$\tan{\beta}$ & 35  &  47    \\
sign($\mu$)   & $+$ & $+$   \\ 
\hline
$\Omega h^2$  & 0.117 &0.113   \\ 
\hline \hline
$m_{\tilde \nu}\;\rm (GeV)$ & 291 & 424  \\
$m_{\chi^0 _1} \;\rm (GeV)$ & 138 & 181 \\
$m_{\chi ^0 _2} \;\rm (GeV)$ & 265 & 350 \\
$m_{\chi ^\pm _1} \;\rm (GeV)$ & 266 & 350 \\
$\Gamma_{\tilde \nu}\;\rm (MeV)$  & 329 & 885 \\
$B_{\chi ^0 _1} \;\rm (GeV)$ & 0.71 & 0.42 \\
$B_{\chi ^0 _2} \;\rm (GeV)$ & 0.09 & 0.19 \\
$B_{\chi ^\pm _1} \;\rm (GeV)$ & 0.20 & 0.39 \\
$\sigma^{peak}_{\tilde \nu}\;$ (nb)  & 39.4 & 19.8  \\
$\sigma^{peak}_{\tilde \nu} \Gamma_{\tilde \nu}\;$ (nb GeV)  & 13.0 & 17.5  \\
\hline
\end{tabular}
\caption{\label{snu-tab} 
Relevant mass spectra, widths, branching fractions and peak cross sections
for the resonant sneutrino scattering, in the two mSUGRA scenarios I, L
defined in Table~\ref{events}.}
\end{table}
The detailed energy spectrum of the hadrons
and leptons coming from the sneutrino decay cascades also depends crucially
on the masses of the particles in each step of the decay chains.
Their study would require
a Monte Carlo treatment of the complete kinematics.
We will not go into this analysis here.

The observed event number is ruled by the product
$\sigma^{peak}_{\tilde \nu } \Gamma_{\tilde \nu }$, that is the peak cross
section (depending on the sneutrino
branching fractions into the initial and final states)
times the resonance width.  

\begin{figure}[H]
\vspace*{1cm}
\centerline{
\mbox{\epsfxsize=13.truecm\epsfysize=13.truecm\epsffile{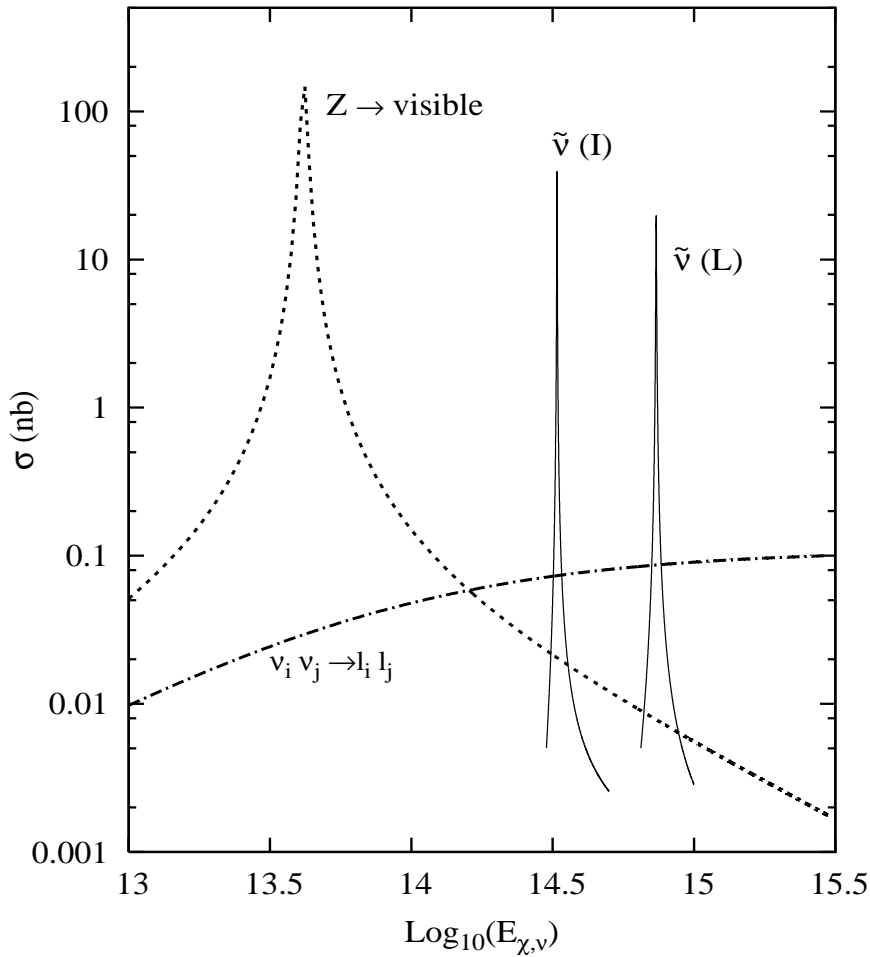}}}
\caption{\label{snu-res}
Cross sections (in nb) for resonant sneutrino production 
and visible decay (solid curves) coming
from UHE neutralino scattering off relic neutrinos in the two mSUGRA
scenarios I, L defined in Table~\ref{events}. 
The dotted line represents the resonant $Z$-boson cross section for the
scattering of UHE neutrinos off relic neutrinos. 
The dot-dashed line refers to  the continuum
cross section for $\nu_i \bar \nu_j^{relic} \rightarrow l_i \bar
l_j$. 
We assumed a relic-neutrino mass $m_\nu= 0.1$ eV. }
\end{figure}
In
Table~\ref{snu-tab}, we present these quantities, along with the relevant mass
spectrum and branching fractions, for the scenarios
I and L, that are the only benchmarks among  the ones presented in  
Table~\ref{events} for which sneutrino {\it visible} decays are allowed by
phase space. In particular, sneutrino decay into the lightest chargino and
second lightest neutralino (that are almost degenerate in mass)
are open. The sneutrino decay into heavier charginos
and $\chi^0_{3,4}$ neutralinos  are not allowed, 
which is important in order not to
deplete too much $B(\tilde \nu\to\chi_1^0 \nu)$ and,
as a consequence, the sneutrino cross section.

In Fig.~\ref{snu-res}, we plot the cross sections for
 the resonant $\tilde
 \nu$ production in the channel in eq.~(\ref{sneut})
 versus the incident neutralino energy, for
 the  scenarios I and L.
 The cross section for the $Z$-production from UHE-$\nu$
 scattering off relic neutrinos
 is also shown versus the incident
 neutrino energy, for comparison. A relic
 neutrino mass of 0.1 eV is assumed in all cases \cite{relic_nu}
 and the resonant incoming UHE
  neutralino energy  is given by  $E^{peak}_{\chi} 
  \simeq (m^2_{\tilde \nu}- m_{\chi}^2)/(2
  m_\nu)$.

Both the sneutrino peak cross section 
and the sneutrino width in scenarios I and L are quite smaller than
(although not very far from, cf. Table~\ref{snu-tab})
the corresponding $Z$ quantities, which are $\sigma^{peak}_Z \simeq 185$ nb
and $\Gamma_Z \simeq 2.5$ GeV.
In particular, one gets $\sigma^{peak}_{\tilde \nu}\cdot \Gamma_{\tilde
\nu} \simeq 13-18$ nb GeV, to be compared with
$\sigma^{peak}_Z\cdot\Gamma_Z \simeq 463$ nb GeV.  

 Of course, the latter
comparison assumes a comparable flux of
incident neutrinos and neutralinos.  
With this hypothesis, the {\it Z-burst} is expected to be dominant on the 
{\it sneutrino-burst}. Nevertheless,
an extra fraction of sneutrino events
could add to the {\it Z-burst}
counting, presumably with a bit less structured energy
distribution. Further in-depth analysis are  needed to
establish the latter point.

Note that, while calculating
the peak cross sections, in both the $Z$ and $\tilde \nu$ cases 
the relic light neutrinos are assumed to be Dirac fermions. In case
of Majorana neutrinos, the peak cross sections 
are enhanced by a factor 2, doubling the number of
the events in each case.

\section{Conclusions}
If elementary-particle interactions  are governed by
supersymmetry, then an important component of the UHECR
 may be given by the lightest stable supersymmetric particle, namely
the lightest neutralino.  For instance, some of the models trying to explain
the UHECR spectrum beyond the GZK cut-off via super-heavy particle
decays, when including  supersymmetry, predict a flux of
UHE neutralinos arriving to the earth. One of the major challenges of
the future cosmic-ray experiments is to verify such predictions.

In this paper, we
computed the event rates for  right selectrons and
right squarks produced on resonance,
when UHE neutralinos scatter off the electrons and quarks in
a detector like IceCube.
We assumed a well-motivated model of supersymmetry, that is minimal
supergravity, and analyzed the signal for moderate
scalar masses, when the resonant scattering is expected to be
dominant on the $t$-channel scattering.

The right scalar sector turns out to be particularly
straightforward to analyze.  One has in general
$B(\tilde {e}_R \rightarrow e\;\chi^0_1)\simeq 1$, and
 $B(\tilde {q}_R \rightarrow q\;\chi^0_1)\simeq 1$. As a consequence,
 after constraining the mSUGRA parameter space
 by negative collider searches and neutralino relic-density bounds,
the relevant
phenomenology depends only on the physical masses of the
resonant right scalar and lightest neutralino.
Any other parameter dependence, such as the one arising from
the neutralino physical compositions and couplings,  drops off.
We performed a detailed study in scenarios inspired
to the post-WMAP mSUGRA benchmarks
of \cite{Battaglia:2003ab}.

For comparison, event rates for the expected resonant 
$W$ signal  in UHE anti-neutrino scattering off electrons in matter
were also presented.

Two different  phenomenological forms for the neutralino fluxes were
assumed to calculate the event rates:
power laws with exponent either $\beta_0=1.5$  or $\beta_0=2$. 
We set the flux normalization on
the basis of present bounds on the UHE neutrino flux from the AGASA
and Fly's Eye experiments (assuming, quite conservatively, that
interaction rates of neutrino and neutralino with matter are of
comparable strength).

For similar
neutrino and neutralino fluxes, in the leptonic channel
the event rates for SUSY particle
resonances are very promising 
(up to a few tens of events per year in IceCube). They
are remarkably greater than the leptonic Glashow $W$ signal
for most of the post-WMAP
benchmarks characterized by moderate scalar masses
($m_{\tilde e_R}\lsim 500$ GeV), 
that fall  either into the `bulk' or into the co-annihilation
region. The corresponding hadronic signal is less structured in 
the energy spectrum, and penalized by the larger resonance
masses ($m_{\tilde q_R}\gsim 600$ GeV). A few events per year in Icecube
are expected in the most promising benchmarks, in the hadronic case.
Signal in the `funnel' and `focus point' benchmarks is depleted
by the large $m_0$ down to less than 0.01 event/year.

 We showed here for the first time the detailed
spectrum of the visible energy released in the $\bar \nu_e e^-$ reaction in
underground detectors, and discussed how
to calibrate the SUSY  signal
through the expected $W$ spectrum.

We also considered the possibility of resonant production of
sneutrinos in the interaction of UHE neutralinos with light relic
neutrinos in the extended Galactic or Local Group  hot neutrino
halo. The visible decay products of the sneutrinos might trigger
air-showers beyond GZK energies, thus mimicking the so called
{\it Z-burst} process. However, even in the most promising
scenarios, we found that the {\it sneutrino-burst} event rate
would be more than one order of magnitude smaller than 
the {\it Z-burst} rate (assuming
similar fluxes for UHE neutrinos and neutralinos).  
Furthermore, the UHE
neutralino energies tuned for the {\it sneutrino-burst}  are usually
higher than for neutrinos in {\it Z-burst}, 
making it less attractive than the standard
{\it Z-burst} model.

In conclusion, we found a well-defined and interesting  window 
of mSUGRA parameters that can be
tested by the next-generation underground detectors. The
corresponding signal may
exceed in some cases the standard neutrino ones. Its imprint may be
disentangled in a realistic way. These high-energy astrophysical traces
could offer a new opportunity to  high-energy astrophysics 
to anticipate  SUSY  discovery at colliders.

\vskip 1.2cm
{\bf Acknowledgments} \\
This work was partially supported by the RTN European Programme
MRTN-CT-2004-503369 (Quest for Unification).

\newcommand{\plb}[3]{{Phys. Lett.} {\bf B#1} #2 (#3)}                  %
\newcommand{\prl}[3]{Phys. Rev. Lett. {\bf #1} #2 (#3) }        %
\newcommand{\rmp}[3]{Rev. Mod.  Phys. {\bf #1} #2 (#3)}             %
\newcommand{\prep}[3]{Phys. Rep. {\bf #1} #2 (#3)}                   %
\newcommand{\rpp}[3]{Rep. Prog. Phys. {\bf #1} #2 (#3)}             %
\newcommand{\prd}[3]{Phys. Rev. {\bf D#1} #2 (#3)}                    %
\newcommand{\np}[3]{Nucl. Phys. {\bf B#1} #2 (#3)}                     %
\newcommand{\npbps}[3]{Nucl. Phys. B (Proc. Suppl.)
           {\bf #1} #2 (#3)} %
\newcommand{\sci}[3]{Science {\bf #1} #2 (#3)}                 %
\newcommand{\zp}[3]{Z.~Phys. C{\bf#1} #2 (#3)}
\newcommand{\epj}[3]{Eur. Phys. J. {\bf C#1} #2 (#3)}
\newcommand{\mpla}[3]{Mod. Phys. Lett. {\bf A#1} #2 (#3)}             %
 \newcommand{\apj}[3]{ Astrophys. J.\/ {\bf #1} #2 (#3)}       %
\newcommand{\jhep}[2]{{Jour. High Energy Phys.\/} {\bf #1} (#2) }%
\newcommand{\astropp}[3]{Astropart. Phys. {\bf #1} #2 (#3)}            %
\newcommand{\ib}[3]{{ ibid.\/} {\bf #1} #2 (#3)}                    %
\newcommand{\nat}[3]{Nature (London) {\bf #1} #2 (#3)}         %
 \newcommand{\app}[3]{{ Acta Phys. Polon.   B\/}{\bf #1} #2 (#3)}%
\newcommand{\nuovocim}[3]{Nuovo Cim. {\bf C#1} #2 (#3)}         %
\newcommand{\yadfiz}[4]{Yad. Fiz. {\bf #1} #2 (#3);             %
Sov. J. Nucl.  Phys. {\bf #1} #3 (#4)]}               %
\newcommand{\jetp}[6]{{Zh. Eksp. Teor. Fiz.\/} {\bf #1} (#3) #2;
           {JETP } {\bf #4} (#6) #5}%
\newcommand{\jetpl}[6]{{\em Pisma Zh. Eksp. Teor. Fiz.\/ } {\bf #1} #2 (#3)
[{\it JETP Lett.\/} {\bf #4} #5 (#6)]}     %
\newcommand{\philt}[3]{Phil. Trans. Roy. Soc. London A {\bf #1} #2
        (#3)}                                                          %
\newcommand{\hepph}[1]{hep--ph/#1}           %
\newcommand{\hepex}[1]{hep--ex/#1}           %
\newcommand{\astro}[1]{astro--ph/#1}         %

\end{document}